\documentclass[aps,pre,reprint,superscriptaddress]{revtex4-2}
\usepackage[latin9]{inputenc}
\usepackage{amstext}
\usepackage{graphicx}

\makeatletter

\newcommand{\lyxmathsym}[1]{\ifmmode\begingroup\def\b@ld{bold}
  \text{\ifx\math@version\b@ld\bfseries\fi#1}\endgroup\else#1\fi}

\providecommand{\tabularnewline}{\\}

\makeatother

\begin{document}
\title{Evolution and Pathogenicity of SARS-CoVs: A Microcanonical Analysis
of Receptor-Binding Motifs}
\author{Rafael B. Frigori}
\email{frigori@utfpr.edu.br}

\affiliation{Universidade Tecnológica Federal do Paraná, Rua Cristo Rei 19, CEP
85902-490, Toledo (PR), Brazil}
\begin{abstract}
The rapid evolution and global impact of coronaviruses, notably SARS-CoV-1
and SARS-CoV-2, underscore the importance of understanding their molecular
mechanisms in detail. This study focuses on the receptor-binding motif
(RBM) within the Spike protein of these viruses, a critical element
for viral entry through interaction with the ACE2 receptor. We investigate
the sequence variations in the RBM across SARS-CoV-1, SARS-CoV-2 and
its early variants of concern (VOCs). Utilizing multicanonical simulations
and microcanonical analysis, we examine how these variations influence
the folding dynamics, thermostability, and solubility of the RBMs.
Our methodology includes calculating the density of states (DoS) to
identify structural phase transitions and assess thermodynamic properties.
Furthermore, we solve the Poisson-Boltzmann equation to model the
solubility of the RBMs in aqueous environments. This methodology is
expected to elucidate structural and functional differences in viral
evolution and pathogenicity, likely improving targeted treatments
and vaccines. 
\end{abstract}
\maketitle

\section{Introduction}

The recent emergence and rapid dissemination of coronaviruses \cite{cui2019origin,song2019sars},
particularly SARS-CoV-1 and SARS-CoV-2, have profoundly impacted global
health, emphasizing the urgent need for a detailed understanding of
their molecular mechanisms, especially those involved in viral entry
and infection \cite{lan2020structure,shang2020cell,jackson2022mechanisms}.
Both SARS-CoV-1, responsible for the Severe Acute Respiratory Syndrome
(SARS) outbreak in 2002-2003, and SARS-CoV-2, which triggered the
COVID-19 pandemic starting in late 2019, belong to the \textit{Coronaviridae}
family. These viruses share a large, enveloped, positive-sense single-stranded
RNA genome, and their origins are closely linked to zoonotic spillovers,
likely from bats to intermediate hosts before reaching humans \cite{xu2020systematic}.
Despite their similarities, the evolutionary paths of these viruses
have led to significant differences in their biological behavior and
impact on human health, making the study of their Spike (S) protein
critical for understanding their pathogenicity and transmissibility
\cite{hatmal2020comprehensive}.

The Spike protein is a trimeric transmembrane glycoprotein that mediates
the attachment of the virus to the host cell surface and the subsequent
fusion of the viral and cellular membranes, which is critical for
viral entry. The receptor-binding domain (RBD) of the Spike protein
is the region that directly interacts with the angiotensin-converting
enzyme 2 (ACE2) receptor on host cells, determining the virus's ability
to infect \cite{li2005structure,tai2020characterization}. Within
the RBD lies receptor-binding motif (RBM), a key subdomain responsible
for the direct interaction with ACE2. This motif is highly conserved
in coronaviruses but exhibits subtle variations between different
strains and variants, influencing the binding affinity and, consequently,
the virus's transmissibility and pathogenicity \cite{yi2020key}.

SARS-CoV-1 and SARS-CoV-2 share a common ancestry but differ significantly
in the structural features of their Spike proteins, particularly in
the RBM. SARS-CoV-1's RBM binds to ACE2 with high affinity, which
correlates with its ability to cause severe respiratory illness \cite{li2003angiotensin,hoffmann2020sars}.
In contrast, SARS-CoV-2's RBM, although it shares structural similarities
with SARS-CoV-1, has evolved to enhance binding affinity to ACE2,
contributing to its higher transmissibility \cite{lan2020structure,hatmal2020comprehensive,walls2020structure,korber2020tracking}.
Notably, early variants of concern of SARS-CoV-2, such as the Beta
(P.1) and Gamma (B.1.351) variants, exhibit mutations within the RBM
that alter key residues involved in ACE2 binding. These mutations,
including the N501Y and E484K, not only enhance the binding affinity
to ACE2 \cite{tegally2021detection,faria2021genomics} but also confer
the ability to evade neutralizing antibodies generated by natural
infection or vaccination \cite{garcia2021multiple,zhou2021evidence}.

Moreover, the impact of these mutations extends beyond immune evasion
\cite{harvey2021sars,carabelli2023sars}. They also influence the
Spike protein's biophysical properties, such as charge distribution,
solubility, and structural stability \cite{ponde2022physicochemical}.
For instance, the E484K mutation, found in both the Beta ($\beta$)
and Gamma ($\gamma$) variants, introduces a positive charge in the
RBM, which can disrupt local electrostatic interactions and modify
the overall stability of the protein \cite{zhang2022surface,WANG2021108035}.
Similarly, mutations like N501Y enhance the Spike protein's structural
stability, which may contribute to more efficient viral entry \cite{ponde2022physicochemical}.
Therefore, the structural and functional differences in the RBM of
those SARS-CoVs underscore the complex relationship between viral
evolution, protein dynamics, and pathogenicity.

From a physical perspective the folding behavior of RBMs, crucial
for their thermostability and Spike interaction potential, can be
rigorously analyzed using multicanonical simulations \cite{berg1991multicanonical,berg1995multicanonical,berg1999introduction,berg2003multicanonical},
which enable precise calculations of the density of states (DoS).
This approach, rooted in D.H.E. Gross's theoretical framework \cite{gross2001microcanonical},
facilitates microcanonical analyzes that uncovers critical phenomena
such as phase transitions \cite{stanley1971phase}, often missed in
simulations using the canonical ensemble \cite{hernandez2008microcanonical}.
RBMs, as protein motifs, qualify as ``small systems'' under this
scenario, where the small number of particles such as the limited
amino acid residues in RBMs allows surface effects and fluctuations
to significantly impact their behavior \cite{bachmann2014thermodynamics}.
Small systems, such as atomic nucleai \cite{chomaz2002nuclear,buyukccizmeci1nuclear},
nanoscale materials \cite{barre2001inequivalence,frigori2010extended,alves2016superstatistics},
and proteins \cite{bachmann2014thermodynamics,frigori2013microcanonical,frigori2014breakout}
as the Spike RBM motifs, exhibit unique thermostatistical properties,
including energy-dependent equilibrium states in the microcanonical
ensemble that lack direct counterparts in the canonical ensemble due
to the nonconcavity of the entropy function \cite{,hernandez2008microcanonical,bachmann2014thermodynamics,barre2001inequivalence}.
This nonconcavity can lead to phenomena like negative heat capacity
and metastable or unstable states, which canonical methods often overlook
\cite{gross2001microcanonical}. Thus, microcanonical analysis is
crucial for accurately capturing these states and understanding the
detailed thermostatistics and functional behavior of RBMs.

In addition to folding dynamics, the solubility of proteins in an
aqueous environment plays a crucial role in their biological function
\cite{levy2006water}. The solubility of these peptides can be described
using the Poisson-Boltzmann equation (PBE) \cite{im1998continuum},
which accounts for the electrostatic interactions between the protein
and the surrounding solvent. By solving the PBE \cite{jo2008charmm,jo2008pbeq}
for each RBM variant, we can gain insights into how changes in the
amino acid sequence influence these protein's domains electrostatic
potential, and consequently, its solubility \cite{frigori2017positive}.
These mutations are also expected to influence the propensity of Spike
to form aggregates or interact with other molecules \cite{IDREES202194,bhardwaj2023amyloidogenic},
with potential consequences for viral infectivity and immune recognition
\cite{yi2020key,greaney2021comprehensive}.

The combination of multicanonical simulations \cite{berg2003multicanonical,meinke2008smmp},
microcanonical analysis \cite{gross2001microcanonical,bachmann2014thermodynamics},
and electrostatic modeling \cite{im1998continuum,frigori2021microcanonical}
provides a powerful methodology for understanding the molecular mechanisms
that underlie the differences between SARS-CoV-1 and various SARS-CoV-2
variants. By delving into the detailed thermostatistics and solubility
properties of the RBM, we aim to uncover the fundamental principles
governing the folding, stability, and function of these critical viral
motifs. This knowledge is essential not only for understanding the
molecular basis of viral evolution and pathogenicity but also for
informing the development of therapeutic strategies and vaccines aimed
at combating current and future coronavirus outbreaks.

The article is organized as follows. In Sections 2 and 3, we describe
the setup for multicanonical simulations used to investigate the folding
dynamics of SARS-CoV RBMs and the microcanonical analysis framework
employed to reveal structural phase transitions in folding, often
missed by canonical ensembles. In Section 4, we present our results,
focusing on the structural phase transitions identified through density
of states calculations and the application of the Poisson-Boltzmann
equation to model RBM solubility. We also discuss the broader implications
of these findings, particularly how RBM mutations influence viral
evolution, pathogenicity, and potential therapeutic strategies. Finally,
in Section 5, we conclude by summarizing our key insights and proposing
directions for future research.

\section{Modeling And Computational Setup}

We perform massive parallel Monte Carlo simulations \cite{zierenberg2014scaling}
within a computational setup previously employed on modeling Amylin,
Insulin, and Amyloid-$\beta$ interactions \cite{frigori2017positive,frigori2021microcanonical,frigori2024insights}.
Here, the focus is on simulating the receptor-binding motifs (RBMs)
of different SARS-CoVs to explore their folding dynamics and thermodynamic
properties. The simulations employ the ECEPP/3 force field and the
implicit solvation model SCH2, as implemented in the SMMP 3.0 package
\cite{meinke2008smmp}, which provides an effective balance between
accuracy and computational efficiency, consistent with Molecular Dynamics
predictions \cite{frigori2017positive,alves2018silico,alves2019synergistic}.
While the absence of explicit solvation is known to introduce phase
transitions at significantly elevated temperatures, this is a minor
drawback that must be accepted as a tradeoff for the high-statistics
data achieved. This approach involves up to 36 million molecular sweeps
across 300 multicanonical (MUCA) recursions, enabling a robust microcanonical
analysis.

It shall be noted that RBMs of SARS-CoV-1 and SARS-CoV-2 has not been
experimentally characterized in isolation, as available structural
data pertain to the RBM embedded within the receptor-binding domain
(RBD) of the Spike protein. The absence of experimental data on the
isolated RBM poses a challenge, as its structural and dynamic properties
may differ significantly from those observed in the native Spike context.
This implies the need of computational modeling to generate initial
structures suitable for simulation studies.

\begin{figure}[ht!]
\centering{}\includegraphics[width=7.5cm]{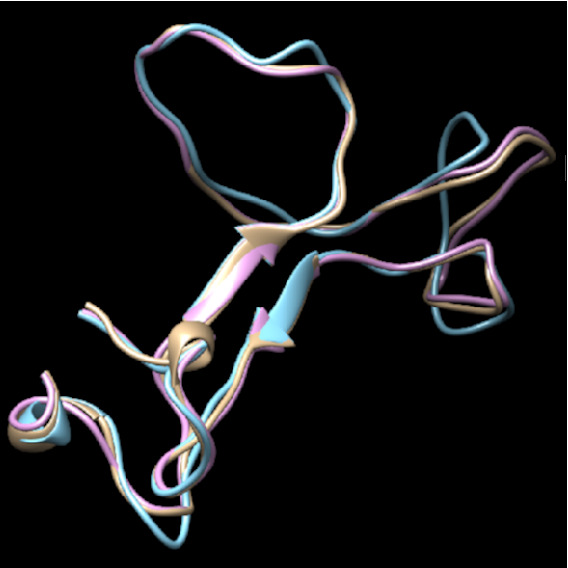} \caption{ RBMs modeled using I-TASSER serve as initial structures for SARS-1
(Blue), SARS-2 WT (Purple), SARS-2 $\beta/\gamma$ (Brown) variants.}\label{RBMs_PDB}
\end{figure}

To begin the simulations, the RBMs for SARS-CoV-1, SARS-CoV-2 and
its VOCs must first be encoded in the one-letter FASTA format, shown
in (Table \ref{RBMs_FASTA}). Noteworthy, the $\beta/\gamma$ VOCs
share the same RBM, whose mutations E484K and N501Y are prone to immunity
evasiveness \cite{harvey2021sars,masson2012viralzone}. Then, these
sequences need to be converted into three-dimensional input structures
(Figure \ref{RBMs_PDB}), for instance using the I-TASSER (Iterative
Threading ASSEmbly Refinement) software for homology modeling \cite{yang2015tasser,zheng2021folding,zhou2022tasser}.
I-TASSER is a robust tool for protein structure and function prediction.
It operates by initially threading the query sequence through a database
of known protein structures to identify appropriate templates. The
software then constructs full-length models via iterative fragment
assembly simulations. This process not only predicts the 3D structure
of the proteins but also offers potential insights into their biological
functions, binding sites, and interactions.

\begin{table*}
\begin{centering}
\begin{tabular}{|c|c|}
\hline 
\textbf{Virus} & \textbf{RBM Sequence (FASTA)}\tabularnewline
\hline 
\hline 
SARS-1 & \texttt{TRNIDATSTGNYNYKYRYLRHGKLRPFERDISNVPFSPDGKPCTPPALNCYWPLNDYGFYTTTGIGYQPY} \tabularnewline
\hline 
SARS-2 (WT) & \texttt{SNNLDSKVGGNYNYLYRLFRKSNLKPFERDISTEIYQAGSTPCNGVEGFNCYFPLQSYGFQPTNGVGYQPY} \tabularnewline
\hline 
SARS-2 ($\beta/\gamma$) & \texttt{SNNLDSKVGGNYNYLYRLFRKSNLKPFERDISTEIYQAGSTPCNGV\underline{\underline{K}}GFNCYFPLQSYGFQPT\underline{\underline{Y}}GVGYQPY} \tabularnewline
\hline 
\end{tabular}
\par\end{centering}
\centering{}\caption{FASTA Sequences of RBMs of the studied SARS-CoVs}\label{RBMs_FASTA}
\end{table*}

The Multicanonical simulations are carried out in a rigid cubic box
with 200$\text{\AA}$ sides, and the Boltzmann constant is set to
$k_{B}=1.987\times10^{-3}$ kcal/mol/K. Although initial molecular
configurations are required to start the simulations, after thermalization,
thermodynamic properties become independent of these initial conditions,
as observed in ergodic Monte Carlo simulations. This allows for a
comprehensive analysis of the folding dynamics and solubility of the
RBMs across a wide temperature range, spanning from the homogeneous
nucleation of ice crystals at 224.8 K to the critical point of water
at 647 K.

\section{Microcanonical Analysis and Solubility Modeling}

Canonical Monte Carlo (MC) simulations are limited in their ability
to explore the full energy landscape of complex systems due to their
reliance on the Boltzmann weight \cite{berg2003multicanonical,hernandez2008microcanonical},
which is defined at a fixed temperature. This constraint often restricts
the utility of re-weighting techniques for extrapolating thermodynamic
quantities across different temperatures \cite{landau2021guide}.
In contrast, Multicanonical (MUCA) simulations \cite{berg1991multicanonical,berg1995multicanonical,berg1999introduction,berg2003multicanonical}
offer a more versatile approach by employing generalized weights that
allow the system to sample a wide range of energy states uniformly.
The MUCA weight is given by

\begin{equation}
\omega_{muca}=\frac{1}{\Omega(E)}=e^{-S(E)}=e^{-\bar{\beta}(E)E+\bar{\alpha}(E)},\label{eq:muca_weights}
\end{equation}
where $\Omega(E)$ denotes the density of states (DoS), $S(E)$ is
the microcanonical entropy, $\bar{\beta}(E)$ is the microcanonical
inverse temperature, and $\bar{\alpha}(E)$ is a dimensionless free
energy term.

The MUCA method comprises two key steps: (1) determining the appropriate
multicanonical weights, usually through an iterative process that
involves histogramming the internal energy, and (2) performing a Markov
Chain Monte Carlo (MCMC) simulation using the weights from the first
step, followed by reweighting to the Gibbs ensemble. In this study,
we focus on microcanonical thermostatistics to analyze the folding
and solubility of the RBM from different SARS-CoV variants. Thus,
our objective is to concentrate on step 1, specifically on computing
the multicanonical entropy

\begin{equation}
S_{muca}(E_{k})=\beta_{k}E_{k}-\alpha_{k},\label{eq:muca_entropy}
\end{equation}
where $\left\{ \beta_{k},\alpha_{k}\right\} $ are piecewise functions
representing the MUCA parameters, which approximate the microcanonical
entropy.

The calculation is implemented through an iterative process \cite{berg1995multicanonical,berg2003multicanonical}
where initial MUCA weights are set to $\omega_{muca}^{(0)}=1$, and
the system undergoes a conventional Metropolis simulation. Data collected
during this simulation is used to construct histograms of the energy
distribution, $H(E)$, which are then used to update the MUCA weights.
The iterative refinement of these weights follows the relation

\begin{equation}
\omega_{muca}^{(n+1)}(E)\equiv e^{-S_{muca}^{(n+1)}(E)}=c\cdot\frac{\omega_{muca}^{(n)}(E)}{H_{muca}^{(n)}(E)},\label{eq:muca_recursion}
\end{equation}
where $c$ is a normalization constant ensuring that $S_{muca}^{(n+1)}(E)$
correctly represents the microcanonical entropy. Traditionally, iterations
aim for histogram flatness; here, we instead monitor for stabilization
of $S_{muca}(E)\times E$ in the whole energetic range, as described
in \cite{frigori2021microcanonical}, which takes $\mathcal{O}\left(5M\right)$
sweeps.

After convergence, the microcanonical thermodynamics \cite{gross2001microcanonical}
of the RBMs folding is automatically analyzed using the PHAST package
\cite{frigori2017phast}. The temperature $T(E)$ as a function of
energy is derived from the microcanonical inverse temperature

\begin{equation}
\beta(E)^{*}=k_{B}\cdot\beta(E)\equiv T^{-1}(E)=\frac{\partial S}{\partial E}\label{eq:microcan_temp}
\end{equation}
where $k_{B}$ is the Boltzmann constant. The microcanonical specific
heat $c_{V}$ is then defined as
\begin{equation}
c_{V}=\frac{dE}{dT}=-\beta{{}^2}/\left(\frac{\partial\beta}{\partial E}\right).\label{eq:microcan_cv}
\end{equation}
Finally, the Helmholtz free energy $F(E)$ can be obtained via a Legendre
transform at a fixed temperature $T_{c}(E)$

\begin{equation}
F(E)=E-S(E)\cdot\left(\frac{\partial S}{\partial E}\right)^{-1}\bigg|_{E=E^{*}(T_{c})}.\label{eq:Helmholtz_energy}
\end{equation}

We thoroughly investigate for potential first-order structural phase
transitions involved in the folding of SARS-CoVs RBMs. In such cases,
microcanonical caloric curves $\beta(E)^{*}\times E$ often exhibit
backbends, or S-bends, which reveal metastable states that remain
hidden in canonical ensemble simulations \cite{gross2001microcanonical,bachmann2014thermodynamics,barre2001inequivalence}.
The pseudo-critical inverse temperatures $T_{c}(E)$ associated with
these transitions are determined using the Maxwell construction across
the energy range $\Delta\tilde{L}=E_{fold}-E_{unfold}$, where $\Delta\tilde{L}$
represents the latent heat of the folding transition, and $E_{fold}$
and $E_{unfold}$ correspond to the internal energies of the folded
and unfolded states, respectively \cite{bachmann2014thermodynamics}.

Furthermore, the solubility of RBMs in water, which is critical for
their biological function, is analyzed using the Poisson-Boltzmann
equation \cite{im1998continuum,frigori2021microcanonical,alves2019synergistic}.
The solvation free energy $\Delta G_{solv}$ is decomposed into nonpolar
($\Delta G_{np}$) and electrostatic ($\Delta G_{elec}$) contributions

\begin{equation}
\Delta G_{solv}=\Delta G_{np}+\Delta G_{elec}.\label{eq:SolvationEnergy}
\end{equation}
The nonpolar contribution, associated with the formation energy of
the molecular cavity, is often calculated using the solvent accessible
surface area (SASA) model which is approximated as $\Delta G_{np}=\gamma\,\text{SASA}$,
where $\gamma$ represents the surface tension, typically treated
as an atom-independent parameter \cite{aguilar2012efficient}. The
electrostatic contribution is given by

\begin{equation}
\Delta G_{elec}=\sum_{k}q_{k}\Delta\varphi_{vs}(r_{k}),\label{eq:EletroSolvEnergy}
\end{equation}
where $q_{k}$ denotes the charge of atom $k$, and $\Delta\varphi_{vs}(r_{k})$
represents the electrostatic potential difference between vacuum and
solvent environments at the position of atom $k$. Accurately solving
the Poisson-Boltzmann equation (PBEQ) to obtain these potentials is
computationally demanding but necessary for precise estimation of
the RBMs solvation free energy. In this study, we use the PBEQ-Solver
\cite{im1998continuum,jo2008charmm,jo2008pbeq} for these calculations,
with the dielectric constants set to $\varepsilon_{s}=78.5$ for the
solvent and $\varepsilon_{p}=2.0$ for the protein, and the salt concentration
set to 150 mM, as in Ref. \cite{frigori2021microcanonical}.

\section{Results and Discussion}

\begin{table*}
\begin{centering}
\begin{tabular}{|c|c|c|c|c|}
\hline 
\textbf{Virus} & \textbf{$T_{fold}\ (\text{K})$} & \textbf{$\Delta G_{solv.}\ (\text{kcal/mol})$} & \textbf{$\Delta\tilde{L}\ (\text{kcal/mol})$} & \textbf{$\Delta F\ (\text{kcal/mol})$}\tabularnewline
\hline 
\hline 
SARS-1 & 402 & -462 & 20.6 & 0.994\tabularnewline
\hline 
SARS-2 (WT) & 541 & -419 & 15.1 & 0.345\tabularnewline
\hline 
SARS-2 ($\beta/\gamma$) & 513 & -465 & --- & ---\tabularnewline
\hline 
\end{tabular}
\par\end{centering}
\centering{}\caption{Thermodynamic results for RBMs of SARS-1, SARS-2 WT and its $\beta/\gamma$
VOCs}\label{Thermo_Results}
\end{table*}

\begin{figure*}[ht!]
\centering{}\includegraphics[width=16cm]{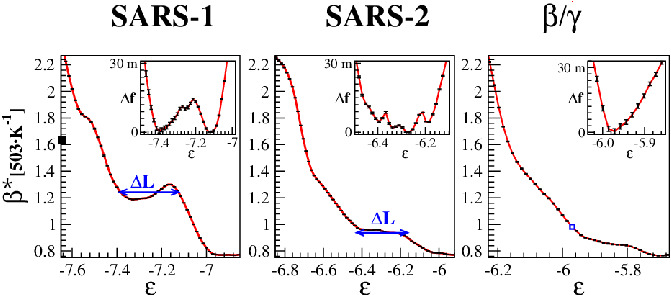}\caption{Microcanonical caloric curve for RBM motifs of SARS-CoVs. The black
square denotes 310K, the bordered square denotes critical transition
temperatures (when existing), $\triangle L$ is the \textit{intensive}
lattent heat of folding obtained by Maxwell's equal-area constructions.
The inserts show the respective \textit{intensive} Helmholtz free
energy $\triangle f$ evaluated at the pseudo- critical temperature
$T_{fold}$. Energy units in $\varepsilon$, $\triangle f$, and $\triangle L$
are normalized to kcal/mol/residues and the physical temperatures
to $503\cdot\beta^{-1}$ K.}\label{Thermo_Fig}
\end{figure*}

\begin{figure*}[ht!]
\centering{}\includegraphics[width=16cm]{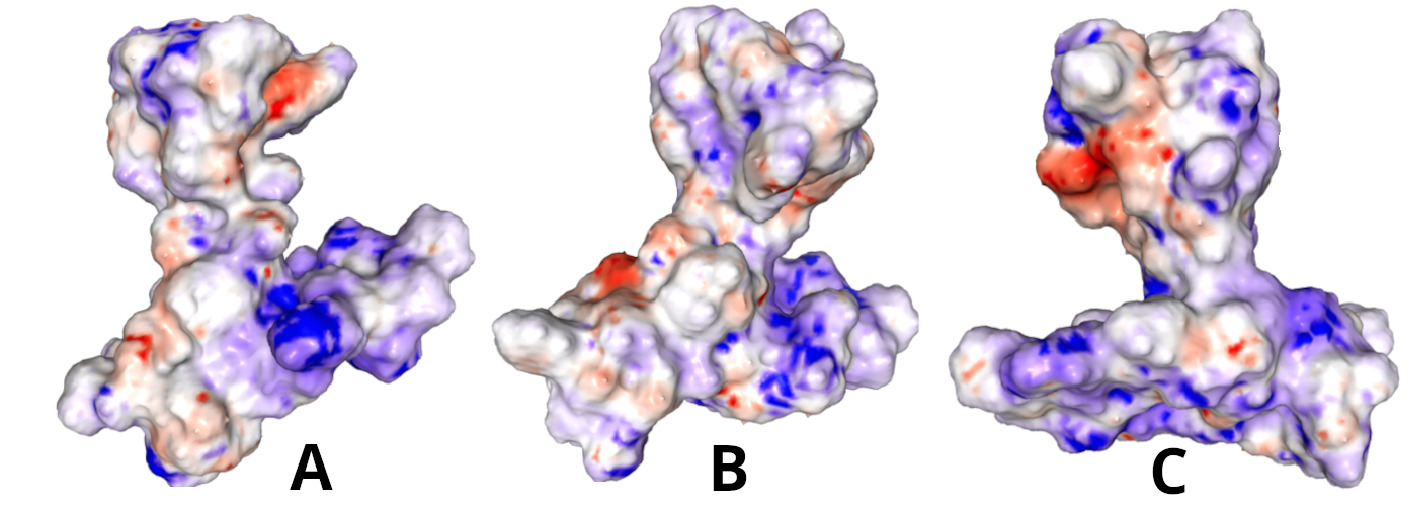}\caption{ Isoelectric surface solutions of PBEQ associated to representative
configurations of RBMs molecular motifs of SARS-1 (A), SARS-2 WT (B)
and SARS-2 $\beta/\gamma$ (C).}\label{PBEQ_Fig}
\end{figure*}

\begin{figure}[t]
\includegraphics[width=8.5cm]{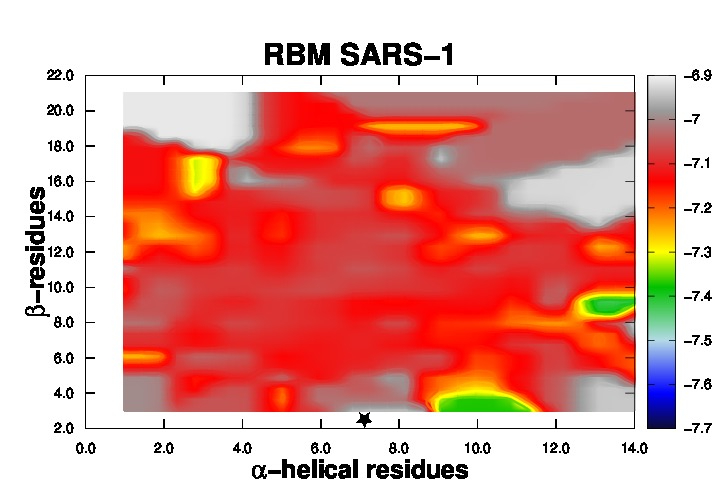}

\includegraphics[width=8.5cm]{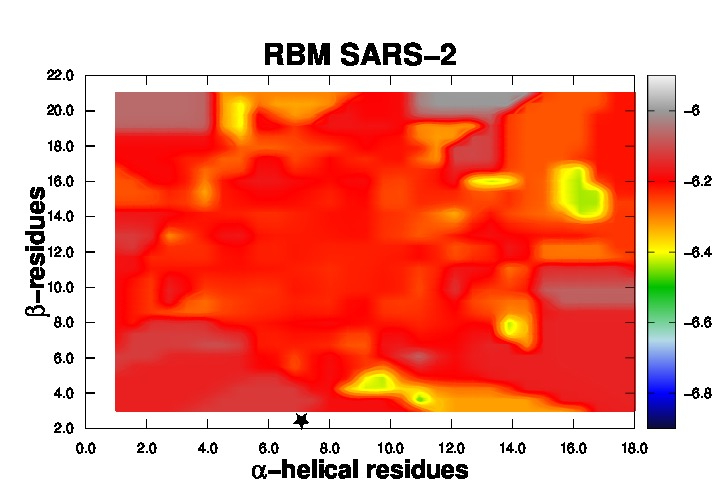}

\includegraphics[width=8.5cm]{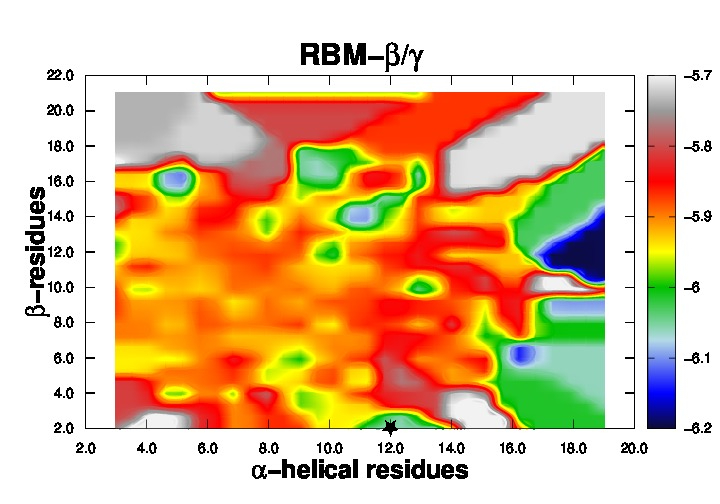}

\caption{Isocontours of the intensive internal energy $\varepsilon$ (kcal/mol/residue)
expressed in terms of the number of residues in $\alpha-$helical
and $\beta-$sheet molecular configurations sampled for RBMs of SARS-1,
SARS-2 WT and SARS-2 $\beta/\gamma$. The star $\left(\star\right)$
shows structural prevalence for I-TASSER initial RBM models.}\label{fig:Isocontours-PhiPsi}
\end{figure}

Parallel simulations were performed to determine the MUCA parameters
$\left\{ \beta_{k},\alpha_{k}\right\} $, which were then utilized
as input for the PHAST package \cite{frigori2017phast}. These parameters
enabled the derivation of the microcanonical thermostatistics governing
the folding structural phase transitions of the RBMs in SARS-CoV-1,
SARS-CoV-2 WT, and its early VOCs $\beta/\gamma$. Thermostatistical
results are summarized in extensive physical units in (Table \ref{Thermo_Results}),
they provide insights into the folding temperatures ($T_{fold}$),
solvation energies ($\Delta G_{solv.}$), latent heats ($\Delta\tilde{L}$)
and free energy barriers ($\Delta F$) that dictate the structural
stability and phase transition behavior of these motifs. The caloric
curves and \textit{intensive} energetic barriers ($\Delta f=\Delta F/$residues)
of the Helmholtz free energy at the (pseudo) critical inverse temperature
($\beta_{c}^{*}$) are depicted in (Figure \ref{Thermo_Results})
as functions of the internal energy per residue ($\varepsilon=E/\text{residues}$).
These plots highlight the existence of regions with canonical metastabilities
during the RBMs folding transitions, where using Maxwell's constructions
we identified the critical $\beta_{c}^{*}$ and the associated \textit{intensive}
latent heats ($\Delta L=\Delta\tilde{L}/$residues) for the SARS-CoV-1
and SARS-CoV-2 WT RBMs. In contrast, the $\beta/\gamma$ variants
exhibit a second-order phase transition, marked by the absence of
latent heat and Helmholtz free energy barriers, indicative of a smooth,
continuous folding process similar to that observed in intrinsically
disordered proteins such as Amylin and Amyloid-$\beta$ \cite{frigori2017positive,frigori2021microcanonical,frigori2024insights}.
The critical $\beta_{c}^{*}\left(\varepsilon_{c}\right)$ of folding
for $\beta/\gamma$ variant is detected by locating the peak of its
$c_{v}\left(\varepsilon\right)\times\varepsilon$ (Eq. \ref{eq:microcan_cv})
--- not shown. Furthermore, (Fig. \ref{PBEQ_Fig}) presents the isoelectric
surfaces of representative RBM configurations for the viruses, sampled
at physiological temperatures ($\sim310$ K), which were used as input
for computing the solvation energies via the PBEQ solver. 

The correlation between thermostatistical outcomes and thermo-structural
features during RBM folding transitions is crucial, especially in
the absence of experimentally resolved RBM structures. Our MUCA simulations,
quantifying residues in $\alpha$-helical and $\beta$-sheet conformations,
illustrate how simulations might provide valuable guidance for future
experimental design. Figure \ref{fig:Isocontours-PhiPsi} shows the
isocontours of internal energy per residue $\left(\varepsilon\right)$
as a function of residues in these configurations. Residue classification
follows the SMMP framework \cite{meinke2008smmp}, using dihedral
angles $\left(\phi,\psi\right)$ with ranges of $\left[-70^{\circ}\pm30^{\circ},-37^{\circ}\pm30^{\circ}\right]$
for helices and $\left[-150^{\circ}\pm30^{\circ},-150^{\circ}\pm30^{\circ}\right]$
for sheets. This figure highlights two notable findings: SARS-CoV-2
is more $\alpha-$helix-prone than SARS-CoV-1, and the E484K and N501Y
substitutions not only shift the folding phase transition to second
order but also significantly enhance the $\alpha-$helical content
in the $\beta/\gamma$ variants' ground state compared to the WT.
This trend, reminiscent of amyloidogenic Amylin isoforms studied using
similar methods \cite{frigori2017phast,alves2018silico,alves2019synergistic},
can be related to the increased pathogenicity of these variants, as
the Spike protein domains are known to induce protein aggregation
under suitable biological conditions \cite{IDREES202194,nystrom2022amyloidogenesis,chang2024sars}.

Next, we delve into the specific thermostatistical and solvation properties
of each RBM, providing a detailed analysis and discussion.

\subsection*{SARS-1: High Solubility and Structural Rigidity as Evolutionary Constraints}

The RBM of SARS-CoV-1 exhibits a pronounced first-order folding transition,
as indicated by a distinct backbend in its caloric curve, presenting
the largest latent heat ($\Delta L=20.6$ kcal/mol), and also highest
Helmholtz free energy barrier among the RBMs studied ($\Delta F=0.994$
kcal/mol). This strong first-order transition points to a stable folded
state, further reinforced by the high folding temperature ($T_{fold}=402$
K). The significant solvation energy of $-462$ kcal/mol reflects
robust interactions with the solvent, resulting in appreciable solubility.

These thermodynamic characteristics suggest that SARS-CoV-1 RBM is
structurally rigid and resistant to mutations. The rigidity and high
structural stability of the RBM limit its ability to undergo significant
conformational changes without compromising its stability, which likely
contributed to the virus's slower evolutionary diversification. The
stringent structural constraints imposed by the RBMs thermodynamic
profile may have hindered the virus's ability to rapidly adapt, leading
to a lower mutation rate compared to SARS-CoV-2. While this rigidity
ensures stable binding to the ACE2 receptor, it also likely reduces
the virus's capacity to evade the host immune system, as the RBM structure
is less flexible in altering epitopes for immune escape.

\subsection*{SARS-2 (WT): Structural Flexibility Facilitates Evolutionary Adaptation}

The RBM of SARS-CoV-2 (RBM-WT) displays a weaker first-order phase
transition compared to SARS-CoV-1, as evidenced by the absence of
backbends in its caloric curve, which displays a latent heat ($\Delta L=15.1$
kcal/mol), and a smaller Helmholtz free energy barrier ($\Delta F=0.345$
kcal/mol). The folding temperature ($T_{fold}=541$ K), derived from
the flat plateau of the caloric curve, reflects a less distinct transition.
This weaker phase transition implies greater structural flexibility
of this RBM, allowing it to adapt to a broader range of structural
perturbations.

Despite this increased flexibility, the solvation energy of the RBM-WT
($-419$ kcal/mol) suggests lower solubility compared to SARS-CoV-1,
potentially driving evolutionary pressure for variants with enhanced
RBM solubility. The structural flexibility of the RBM-WT is critical
for the viruses evolutionary adaptability, enabling it to tolerate
mutations while maintaining effective ACE2 binding. This adaptability
promotes rapid mutation and evolution, which are vital for immune
evasion and sustained transmission. Additionally, the RBM-WT's lower
solubility indicates a potential for further mutations that could
enhance solubility while allowing structural modifications that alter
antibody recognition sites, thereby improving immune evasion without
significantly compromising ACE2 binding.

\subsection*{SARS-2 VOCs: Enhanced Pathogenicity and Immune Evasion}

The RBM of $\beta/\gamma$ variants, characterized by the E484K and
N501Y mutations, present a particular thermodynamic profile, undergoing
a second-order phase transition. Unlike the first-order transitions
observed in the RBMs of SARS-CoV-1 and SARS-CoV-2 WT, these variants
do not exhibit latent heat of folding or Helmholtz free energy barriers,
features typically also observed in fast folders as Intrinsically
Disordered Proteins (IDPs) such as Amylin and Amyloid-$\beta$ \cite{frigori2017positive,frigori2021microcanonical}.
The absence of these thermodynamic signatures indicates smoother,
continuous transitions between conformational states, enhancing the
flexibility and adaptability of the RBM. The increased solvation energy
of $-465$ kcal/mol suggests that these variants have a higher solubility
profile, similar to SARS-CoV-1, which might have emerged by evolutionary
pressure as a mechanism to improve immune evasion of WT antibodies.
The E484K mutation introduces a positively charged lysine at a key
binding interface, altering electrostatic interactions and facilitating
immune evasion \cite{garcia2021multiple,zhou2021evidence,wrobel2022evolution}.
The N501Y mutation replaces asparagine with tyrosine, which introduces
a bulkier, hydrophobic side chain capable of forming stronger $\pi$-stacking
interactions with ACE2 residues, particularly Y41 and K353 \cite{han2021molecular}.
This enhances the affinity for ACE2 binding, stabilizing the RBM-ACE2
interface and further contributing to immune escape by promoting tighter
receptor engagement. Together, these mutations synergistically exploit
structural flexibility and solvent interactions. Therefore, our data
confirm experimental findings in which such mutations boost immune
evasion and ACE2 binding in these variants \cite{garcia2021multiple,zhou2021evidence,harvey2021sars,masson2012viralzone},
highlighting the underlying role of thermodynamic principles guiding
viral evolution.

\section{Conclusion}

This study provides a detailed microcanonical analysis of the RBMs
in SARS-CoV-1, SARS-CoV-2 WT, and the $\beta/\gamma$ VOCs, offering
critical insights into how thermodynamic properties influence viral
evolution and pathogenicity. By utilizing the ECEPP/3 force field
with implicit solvation, the analysis effectively balances computational
efficiency with molecular accuracy, enabling robust data collection
despite trade-offs such as elevated structural transition temperatures.
The results highlight distinct thermodynamic and structural differences
across these RBMs: SARS-CoV-1\textquoteright s RBM, with a strong
first-order phase transition and high solvation energy, exhibits structural
rigidity that limits mutational flexibility and impedes immune evasion,
reducing transmissibility. In contrast, the SARS-CoV-2 WT RBM shows
greater thermodynamic flexibility, marked by a weaker first-order
phase transition and reduced solvation energy, which promotes a broader
mutational spectrum, facilitating the emergence of variants with enhanced
immune evasion and transmissibility under selective pressures. The
$\beta/\gamma$ VOCs, including the E484K and N501Y mutations, illustrate
these evolutionary dynamics, increasing RBM solvation energy and shifting
folding transitions to second-order, similar to intrinsically disordered
proteins. This structural adaptation supports improved immune evasion
and enhanced ACE2 receptor binding. Furthermore, the analysis reveals
that SARS-CoV-2 RBMs exhibit a stronger propensity for $\alpha$-helix
formation compared to SARS-CoV-1, with the $\beta/\gamma$ variants
showing significant increases in $\alpha$-helical content in their
ground states. This trend, reminiscent of amyloidogenic proteins like
Amylin, may be linked to the increased pathogenicity of these variants,
as the Spike protein domains are known to promote aggregation under
certain biological conditions. These findings underscore the importance
of ongoing surveillance and targeted interventions to mitigate the
risks posed by emerging variants, as these thermodynamic properties
suggest continued potential for viral evolution under selective pressures.

\section*{Acknowledgment}

The Brazilian National Laboratory for Scientific Computing (LNCC)
is acknowledged by supercomputing time granted at the Santos Dumont
facility under project PHAST2. This study was financed in part by
the Coordenação de Aperfeiçoamento de Pessoal de Ní­vel Superior -
Brasil (CAPES) - Finance Code 001. \textit{This article is a tribute
to the memory of my esteemed colleague and friend, Dr. Marcelo Fernandes,
who passed away prematurely due to COVID-19}.

\bibliographystyle{aapmrev4-2}
\bibliography{references}

\end{document}